# Superconductivity at 31 K in "111" type iron arsenide superconductor $Na_{1-x}FeAs$ induced by pressure

S. J. Zhang[1], X. C. Wang[1], Q. Q. Liu[1], Y.X. Lv[1], X. H. Yu[1,2], Z. J. Lin[2].,Y.S. Zhao[2], L. Wang[3], Y. Ding[3], H. K. Mao[3], C. Q. Jin[1]

[1]Institute of Physics, Chinese Academy of Sciences, Beijing100190, China

[2] LANSCE, Los Alamos National Laboratory, Los Alamos, NM 87545, USA

[3] HPCAT, Geophysical Laboratory, Carnegie Institution of Washington, Argonne, USA

**Abstract –** The effect of pressure on superconductivity of "111" type $Na_{1-x}FeAs$ is investigated through temperature dependent electrical resistance measurement in a diamond anvil cell. The superconducting transition temperature ($T_c$) increases from 26 K to a maximum 31 K as the pressure increases from ambient to 3 GPa. Further increasing pressure suppresses $T_c$ drastically. The behavior of pressure tuned $T_c$ in $Na_{1-x}FeAs$ is much different from that in $Li_xFeAs$, although they have the same $Cu_2Sb$ type structure.

**Introduction**

The discovery of superconductivity at 26 K in $LaO_{1-x}F_xFeAs$ [1] by Prof. Hosono Lab opens a new era for high temperature superconductor research. The $T_c$ of this material was further raised to 55 K at ambient pressure by replacing La with other rare earth ions with smaller radius [2, 3]. The transition temperature is only second to the high $T_c$ cuprate superconductors. Subsequently, superconductivity with a relative high transition temperature $T_c$ was found in several other "1111" type iron pnictide compounds [4-6]. Other than the aforementioned "1111" type compounds (REFeAsO, RE=rare earth), the "122" type $BaFe_2As_2$ with a tetragonal $ThCr_2Si_2$-type structure was found to be superconducting at 38 K by hole doping [7]. More recently, we found $Li_xFeAs$, a "111" type iron arsenide compound with a $Cu_2Sb$ type tetragonal structure, to be superconducting with a transition temperature of 18 K [8]. With element substitution, the isostructural $Na_{1-x}FeAs$ was also found to be superconducting with $T_c$ 9～26 K [9，10，11] at ambient pressure. Different from the "1111" type or "122" type compounds, the spin density wave (SDW) transition seems absent in $Li_xFeAs$, as derived from experimental observations [8, 12, 13]. The pressure-tuned superconductivity has been investigated for many iron arsenide compounds to enhance the superconductivity transition temperature as well as to understand the mechanism of superconductivity in iron arsenide superconductors. It was found that pressure enhanced its Tc to 43K right after the discovery of superconductivity in $LaO_{1-x}F_xFeAs$ [14]. For some "1111" type and "122" type parent compounds, the superconductivity can be initiated by pressure and the $T_c$ can be pushed to a maximum value with initial compression, then the $T_c$ decreases at higher pressure region, *e.g.* for LaFeAsO [15] and $AFe_2As_2$ (A= Sr, Ba) [16, 17]. For doped $REFeAsO_{1-x}$ with smaller RE ion radius, the $T_c$ is suppressed monotonously with increasing pressure [18, 19]. The $T_c$ is also suppressed linearly with pressure for the "111" type $Li_xFeAs$ [20, 21, 22]. More recently, α-FeSe, with a structure composed of anti-PbO-type FeSe layers, was found to exhibit superconductivity at about 8 K at ambient pressure [23] and showed a dramatic enhancement of $T_c$ by applying low pressure [24, 25]. Pressure is therefore a very important parameter to study iron pnictide superconductors. Here we report the pressure effects on superconductivity of "111" type $Na_{1-x}FeAs$. The results are compared with those for the isostructural superconductor $Li_xFeAs$.

**Experimental details**

The $Na_{1-x}FeAs$ compound used in the experiment was synthesized by solid state reaction method using $Na_3As$, Fe and As as starting materials following the method described in ref.8. Considering volatility loss of Na in the sintering process, the $Na_3As$ precursor powder, Fe and As powder were mixed according to mole ratio of Na:Fe:As = 1.2:1:1 that would give rise to a pure "111" type structural sample with some sodium vacancies. The mixture was pressed into pellet and wrapped with Ta foil in a glove box with high purity argon atmosphere. The pellet wrapped by the Ta foil was then sealed under vacuum in a quartz tube and sintered at $800^{o}C$ for 20 hours. The resulting sample was characterized by x-ray powder diffraction with a Mac Science diffractometer. Diffraction data was collected with 0.02° and 15 s /step. The composition of the sample was analyzed using an inductively coupled plasma (ICP) spectrometer. The results were Na:Fe:As = 0.86:1:1 indicating there exists vacancies at sodium site that contribute to generate carriers [9, 11].

The pressure-induced evolution of Superconducting transition in $Na_{1-x}FeAs$ was investigated by four-probe electrical resistance measurement methods in a diamond anvil cell (DAC) at variant pressures. In our experiment, pressure was generated by a pair of diamonds with 600-μm-diameter culet. The stainless steel gasket was pre indented from 250 μm to ~40 μm thickness before drilled a 250 μm hole in the center that serves as the sample chamber. The sample hole was covered with a thin layer of cubic boron nitride (BN) for electrical insulation between the gasket and the electrodes. Gold wire of 18μm diameter was used as electrode leads. The $Na_{1-x}FeAs$ sample was laid in the center of the four electrodes with pressure media around. The sample size was about 200 μm x 100 μm x 20 μm. MgO fine powder was used as the pressure-transmitting medium in the experiment. The pressure was measured at room temperature by the ruby fluorescence method before and after each temperature cooling down. The highly hygroscopic nature of the $Na_{1-x}FeAs$ sample makes it very difficult to get a good electric contact when preparing the electrodes in air. We prepared the electrodes as fast as possible (less than 30 Min) to reduce the reaction time of the sample surface with water.

**Results and discussion**

The diffraction pattern of the $Na_{1-x}FeAs$ sample can be indexed by the Cu2Sb type structure with P4/nmm symmetry, as shown in Fig.1, isostructural with $Li_xFeAs$ [8]. Fig.2 shows the electric resistance of $Na_{1-x}FeAs$ as a function of temperature at different pressures up to 8GPa. It also shows that the superconducting transition becomes sharper with initial increasing pressure and gets broader at higher pressures. This behavior is probably related to the reactive nature of $Na_{1-x}FeAs$ sample. The broader transition width at ambient pressure is also observed by other groups (Ref. 11). The increased pressure gradient at higher pressure region causes the broadening of the resistance transition.

The $T_c$ values at variant pressures are determined from the initial deviation from the extrapolated line of the R-T curve as shown in Fig.3. The pressure dependence of $T_c$ of the $Na_{1-x}FeAs$ sample is shown in Fig.4. It is noteworthy that $T_c$ increases as the pressure increases from ambient to 3 GPa, followed by quick decrease at higher pressure. The maximum $T_c$ of 31 K is observed at about 3 GPa. The effect of pressure on $T_c$ for $Na_{1-x}FeAs$ is compared with other iron arsenide superconductors as shown in Fig.5. Two types of behaviors of the superconducting transition evolution with pressure were observed in these iron arsenide superconductors. For the first type, the $T_c$ is enhanced or induced by initial compression, and then decreases at higher pressure. This behavior is observed in $LaFeAsO_{1-x}F_x$ [14, 15], $AFe_2As_2$ (A= Sr, Ba) [16, 17] and $Na_{1-x}FeAs$ in present work. The second type is that the $T_c$ is suppressed by applied pressure. This is observed in $Li_xFeAs$ [20, 21, 22] or in $REFeAsO_{1-x}$ where RE is rare earth elements with smaller ion radius than La [18, 19].

The crystal chemistry parameters such as bond distance or bond angle are critical to superconducting transition temperatures for the iron arsenide superconductors. Our primary experiments of high pressure synchrotron x ray diffractions indicated that $Na_{1-x}FeAs$ keeps stable at least up to 20 GPa. Therefore the superconductivity evolution observed in the present work is merely caused by the changes of electronic structure at high pressure. There are two ways to generate or initiate superconductivity in iron based superconducting systems: chemical doping or applied pressure. Both of them can result in the change of electronic structure through inducing carriers into [FeAs] layers. Here the $FeAs_4$ tetrahedron geometry is considered crucial to determine superconducting transition temperature. According to experimental results in Ref.26, the ideal As−Fe−As bond angle of $\alpha = \beta = 109.47°$ corresponds to the highest $T_c$ of the "1111" system. The results suggest that the change of $T_c$ with chemical doping is much related to the structural distortion from the ideal $FeAs_4$ tetrahedron. Furthermore, the high-pressure angle-dispersive X-ray diffraction experiments on $NdO_{0.88}F_{0.12}FeAs$ show that the As−Fe−As bond angles gradually deviate from ideal tetrahedron values with applied pressure [27]. It is consistent with the experimental result that the $T_c$ is suppressed with compression in $NdFeAsO_{1-x.}$ Not only the deviation of As-Fe-As bond angle from the ideal $FeAs_4$ tetrahedron results in the change of density of states (DOS) at Fermi surface, but also the decreasing Fe-Fe distance will broaden the energy band width that usually gives rise to the decrease of its DOS at Fermi surface. It is also confirmed by calculations that the tuning Fe-As distance modifies both the DOS near Fermi level and the magnetic moment [28]. Therefore the change of intraplanar Fe-Fe distance, Fe-As distance and $FeAs_4$ tetrahedron distortion with pressure will work together leading to the evolution of superconducting transition temperature. Further more systematic synchrotron x ray diffraction experiments for $Na_{1-x}FeAs$ at high pressure is needed in order to get quantitative understanding about the chemical bonding length or angle based on Rietveld refinements.

In summary, the superconducting transition temperature of "111" type $Na_{1-x}FeAs$ was enhanced to 31 K at 3GPa , reaching the record high $T_c$ in the "111" system. The pressure effects on $T_c$ for isostructural $Li_xFeAs$ & $Na_{1-x}FeAs$ are different:

pressure suppresses $T_c$ for $Li_xFeAs$ while enhances $T_c$ for $Na_{1-x}FeAs$. This is assumed to relate with the pressure tuned geometric evolution .

## Acknowledgments:

This work was partially supported by NSF & MOST of China. The work involving Los Alamos National Laboratory was supported by LANL/LDRD program and LANL is operated by Los Alamos National Security LLC under DOE contract DEAC52-06NA25396. HPCAT is supported by DOE-BES, DOE-NNSA, NSF, and the W.M. Keck Foundation. APS is supported by DOE-BES, under Contract No. DE-AC02-06CH11357. This material is also based upon work supported as part of the EFree, an Energy Frontier Research Center funded by the U.S. Department of Energy, Office of Science, Office of Basic Energy Sciences under Award Number DE-SC0001057.

**REFERENCES**

[1] Y. Kamihara, T. Watanabe, M. Hirano, and H. Hosono, **J. Am. Chem. Soc.130**, 3296(2008).

[2] X.H. Chen, T. Wu, G. Wu, R. H. Liu, H. Chen, and D. F.Fang, **Nature 453**, 761 (2008).

[3] Z.-A. Ren, J. Yang,W. Lu,W. Yi, X.-L. Shen, Z.-C. Li, G.-C. Che, X.-L. Dong, L.-L. Sun, F. Zhou, and Z.-X. Zhao,**Europhys. Lett. 82**, 57002 (2008).

[4] G.F. Chen, Z. Li, D. Wu, G. Li, W. Z. Hu, J. Dong, P.Zheng, J.L. Luo, and N.L. Wang, **Phys. Rev. Lett. 100**,247002 (2008).

[5] C. Wang, L.J. Li, S. Chi, Z.W. Zhu, Z. Ren, Y.K. Li, Y.T.Wang, X. Lin, Y.K. Luo, S. Jiang, X.F. Xu, G.H. Cao, and Z.A. Xu, **Europhys. Lett. 83**, 67006 (2008).

[6] P. Cheng, B. Shen, G. Mu, X.Y. Zhu, F. Han, B. Zeng and H.H. Wen, **Europhys. Lett. 85**, 67003 (2009).

[7] M. Rotter, M. Tegel, and D. Johrendt, **Phys. Rev. Lett. 101**, 107006 (2008).

[8] X.C.Wang, Q.Q. Liu, Y.X. Lv, W.B. Gao, L.X.Yang, R.C. Yu, F.Y.Li, C.Q. Jin, **Solid State Communications 148,** 538(2008)

[9] X. C. Wang, Y. X. Lv et al, **Frontiers of Physics in China** (submitted)

[10] G. F. Chen, W. Z. Hu, J. L. Luo, and N. L. Wang, **Phys. Rev. Lett. 102**, 227004 (2009).

[11] Dinah R. Parker, Michael J. Pitcher, Peter J. Baker, Isabel Franke, Tom Lancaster, Stephen J. Blundell and Simon J. Clarke, **Chem. Commun.,** 2189 (2009)

[12] Michael J. Pitcher, Dinah R. Parker, Paul Adamson, Sebastian J. C. Herkelrath, Andrew T. Boothroyd, Simon J. Clarke, **Chem. Commun.,** 5918(2008).

[13] Joshua H. Tapp, Zhongjia Tang, Bing Lv, Kalyan Sasmal, Bernd Lorenz, Paul C.W. Chu, Arnold M. Guloy, **Phys. Rev. B 78**, 060505(2008).

[14] H. Takahashi, K. Igawa, K. Arii, Y. Kamihara, M. Hirano, H. Hosono, **Nature 453**, 376(2008).

[15] H. Okada, K. Igawa, H. Takahashi, Y. Kamihara, M. Hirano, H. Hosono, K. Matsubayashi, and Y. Uwatoko, **J. Phys. Soc. Jpn. 77**, 113712(2008)

[16] Awadhesh Mani, Nilotpal Ghosh, S. Paulraj, A. Bharathi, and C. S. Sundar, **arXiv: 0903.4236** (2009).

[17] K. Igawa, H. Okada, H. Takahashi, S. Matsuishi, Y. Kamihara, M. Hirano, H. Hosono, K. Matsubayashi, Y. Uwatoko, **J. Phys. Soc. Jpn.78**, 025001(2009).

[18] N. Takeshita, A. Iyo, H. Eisaki, H. Kito, and T. Ito, **J. Phys. Soc. Jap 77**, 075003(2008).

[19] W. Yi, L.L. Sun, Z. Ren, W. Lu, X.L. Dong, H.J. Zhang, X. Dai, Z. Fang, Z.C. Li, G.G. Che, J. Yang, X.L. Shen, F. Zhou, Z.X. Zhao, **EPL 83,** 57002(2008).

[20] S. J. Zhang, X. C. Wang, R. Sammynaiken, J. S. Tse, L. X. Yang, Z. Li, Q. Q. Liu, S. Desgreniers, Y. Yao, H. Z. Liu, and C. Q. Jin, **Phys. Rev. B. 80,** 014506 (2009).

[21]Melissa Gooch, Bing Lv, Joshua H. Tapp, Zhongjia Tang, Bernd Lorenz, Arnold M. Guloy and Paul C.W. Chu, **EPL 85,** 27005(2009)

[22] Masaki Mito, Michael J. Pitcher, Wilson Crichton, Gaston Garbarino, Peter J.Baker, Stephen J. Blundell, Paul Adamson, Dinah R. Parker, and Simon J. Clarke, **J. Am. Chem. Soc.131**, 2986(2009)

[23] F. C. Hsu, J. Y. Luo, K. W. Yeh, T. K. Chen, T. W. Huang, P. M. Wu, Y. C. Lee, Y. L. Huang, Y. Y. Chu, D. C. Yan and M. K. Wu, **PNAS 105**, 14262(2008)

[24]Yoshikazu Mizuguchi, Fumiaki Tomioka, Shunsuke Tsuda, Takahide Yamaguchi, and Yoshihiko Takano, **Appl. Phys. Lett. 93**, 152505 (2008)

[25] S. Margadonna, Y. Takabayashi, Y. Ohishi, Y. Mizuguchi, Y.Takano, T. Kagayama, T. Nakagawa, M. Takata, and K. Prassides, **arXiv:0903.2204v1**(2009)

[26 ] C. H. Lee, A. Iyo, H. Eisaki, H. Kito, M. T. Fernandez-Diaz, T. Ito, K. Kihou, H. Matsuhata, M. Braden and K. Yamada, **J. Phys. Soc. Jpn. 77**, 083704(2008).

[27 ] J. G. Zhao, L. H. Wang, D. W. Dong, Z. G. Liu, H. Z. Liu, G. F. Chen, D. Wu, J. L. Luo, N. L. Wang, Y. Yu, C. Q. Jin, and Q. Z. Guo, **JACS 130**, 13828 (2008).

[28] S. Lebègue, Z. P. Yin and W. E. Pickett, **New Journal of Physics 11**, 025004(2009).

**Figure captions**

Fig.1: (color online) X-ray diffraction spectrum of the polycrystalline sample with nominal composition $Na_{1.2}FeAs$. The $Na_{1-x}FeAs$ crystallizes into Cu2Sb type structure with space group P4/nmm. The large (red) spheres are Na atoms, the medium (blue) spheres are As atoms, and the small (yellow) spheres are Fe atoms.

Fig.2: (color online) The temperature dependence of resistance for $Na_{1-x}FeAs$ superconductor at variant pressures from 0 GPa to 8 GPa; The insert shows the details of superconductivity transition with pressure from 3 GPa to 6.5 GPa.

Fig.3: (color online) The R-T curve at 6.5 GPa. The insert shows the definition of $T_c$: the extrapolated line is drawn through the resistivity curve in the normal state just above $T_c$. The $T_c$ is determined from the initial deviation from this line.

Fig.4: The $T_c$-$P$ phase diagram of $Na_{1-x}FeAs$ obtained from resistance measurements. Points are experimental data, while the lines are polynominal fit to the experimental data.

Fig.5: (color online) The comparison of $T_c$-$P$ phase diagrams of iron arsenide superconductors. All points are experimental data with reference indicated.

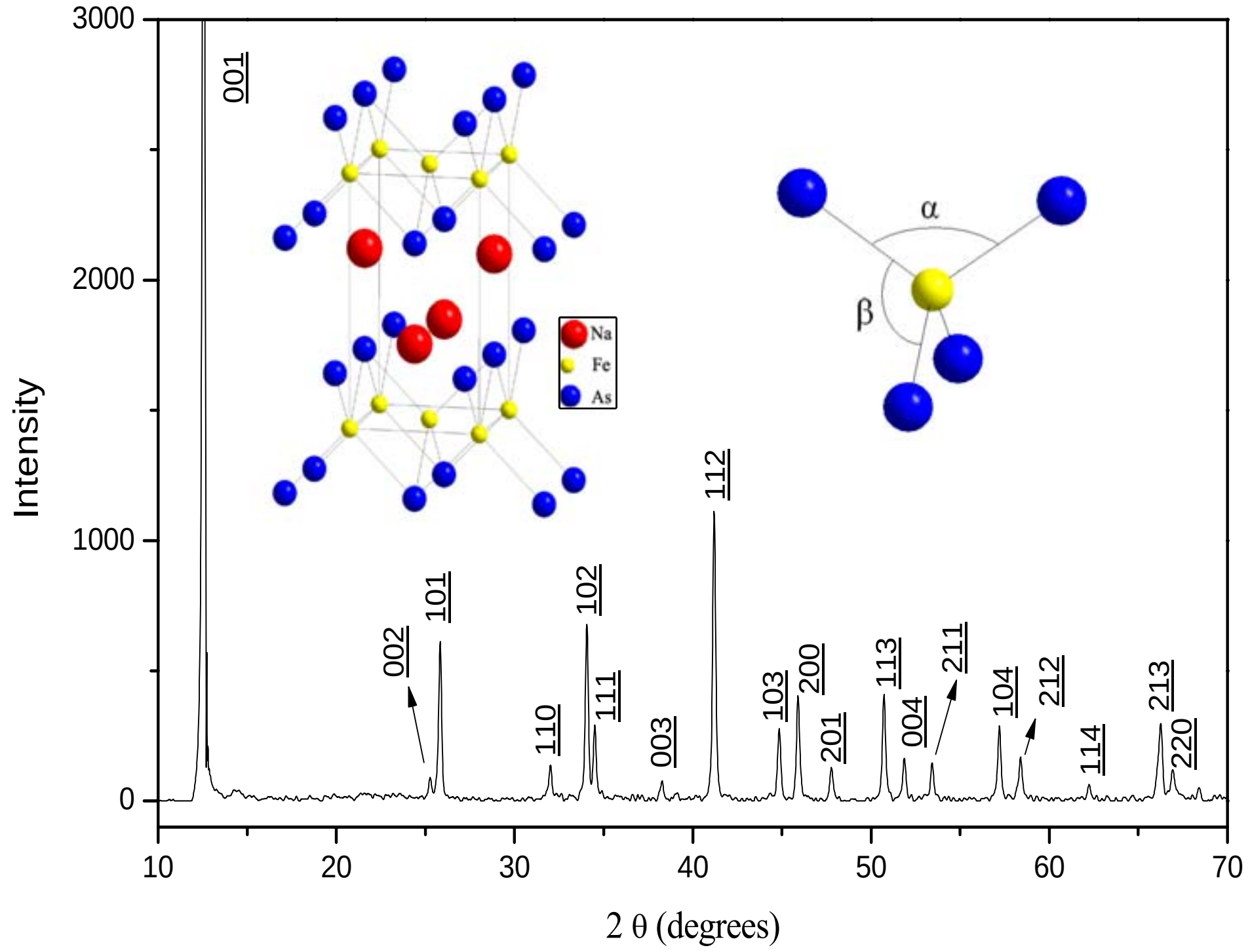

Fig.1

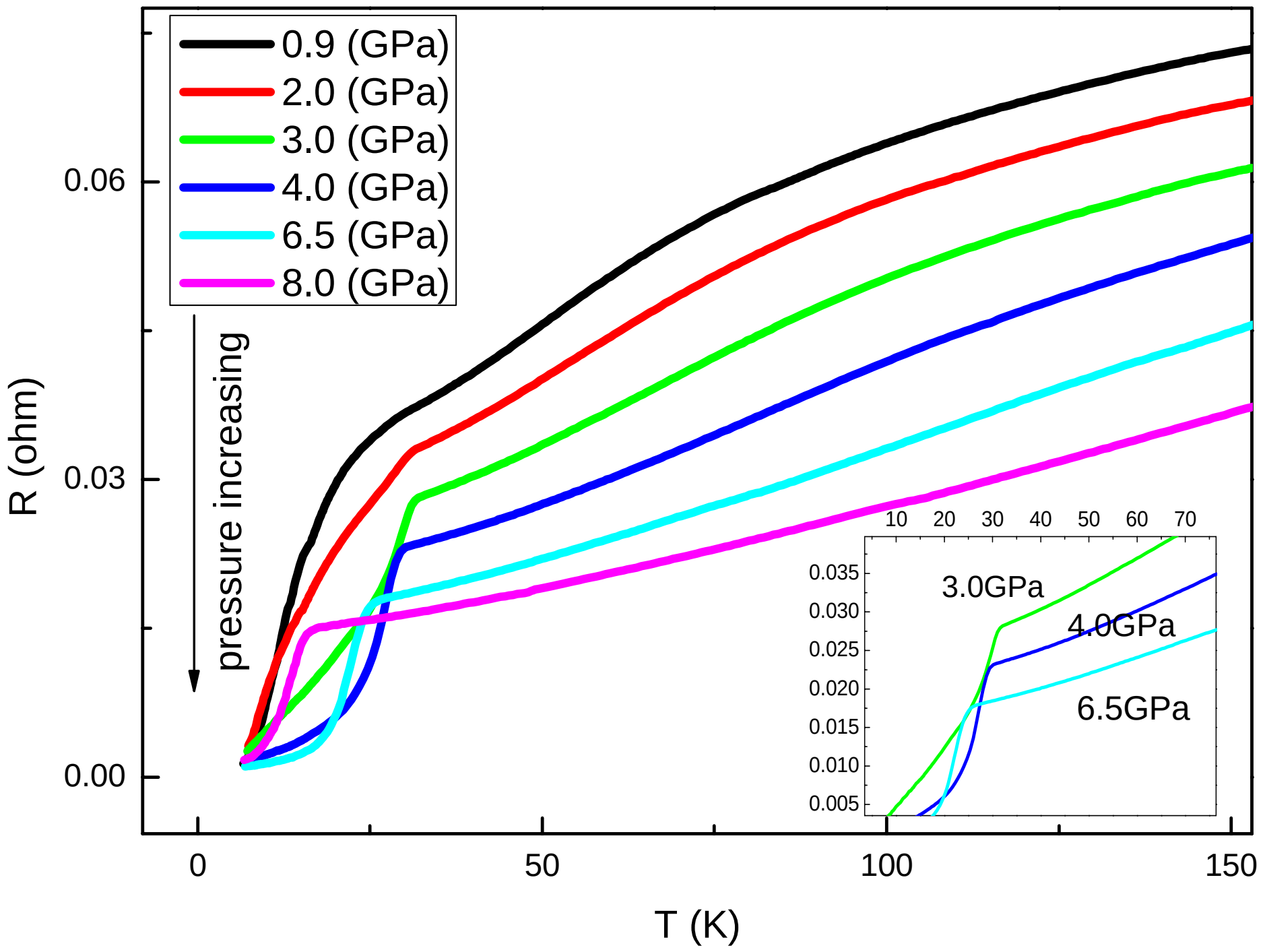

Fig.2

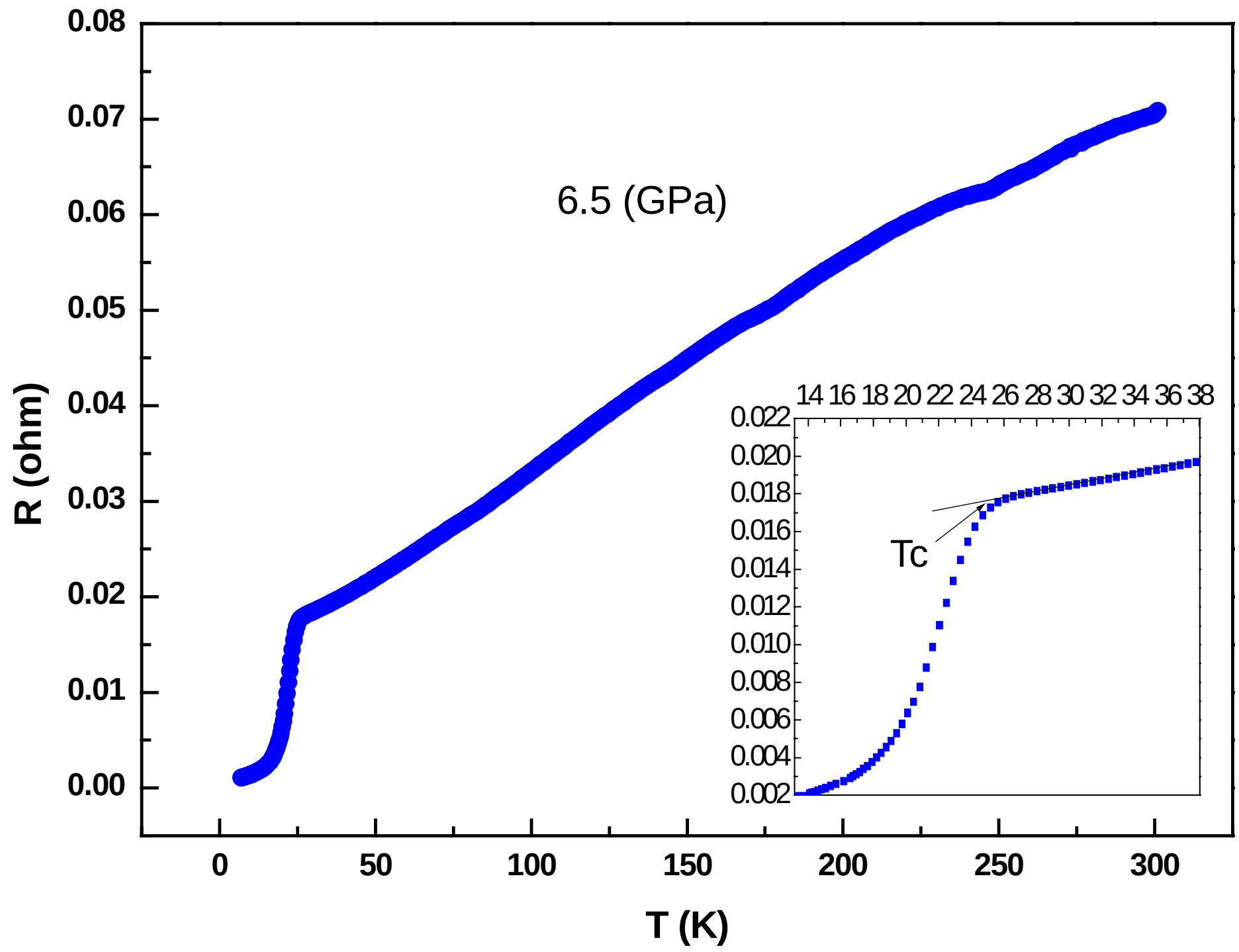

0.08
0.07
0.06
0.05
0.04
0.03
0.02
0.01
0.00
R (ohm)
6.5 (GPa)
0
50
100
150
200
250
300
T (K)
14 16 18 20 22 24 26 28 30 32 34 36 38
0.022
0.020
0.018
0.016
0.014
0.012
0.010
0.008
0.006
0.004
0.002
Tc

Fig.3

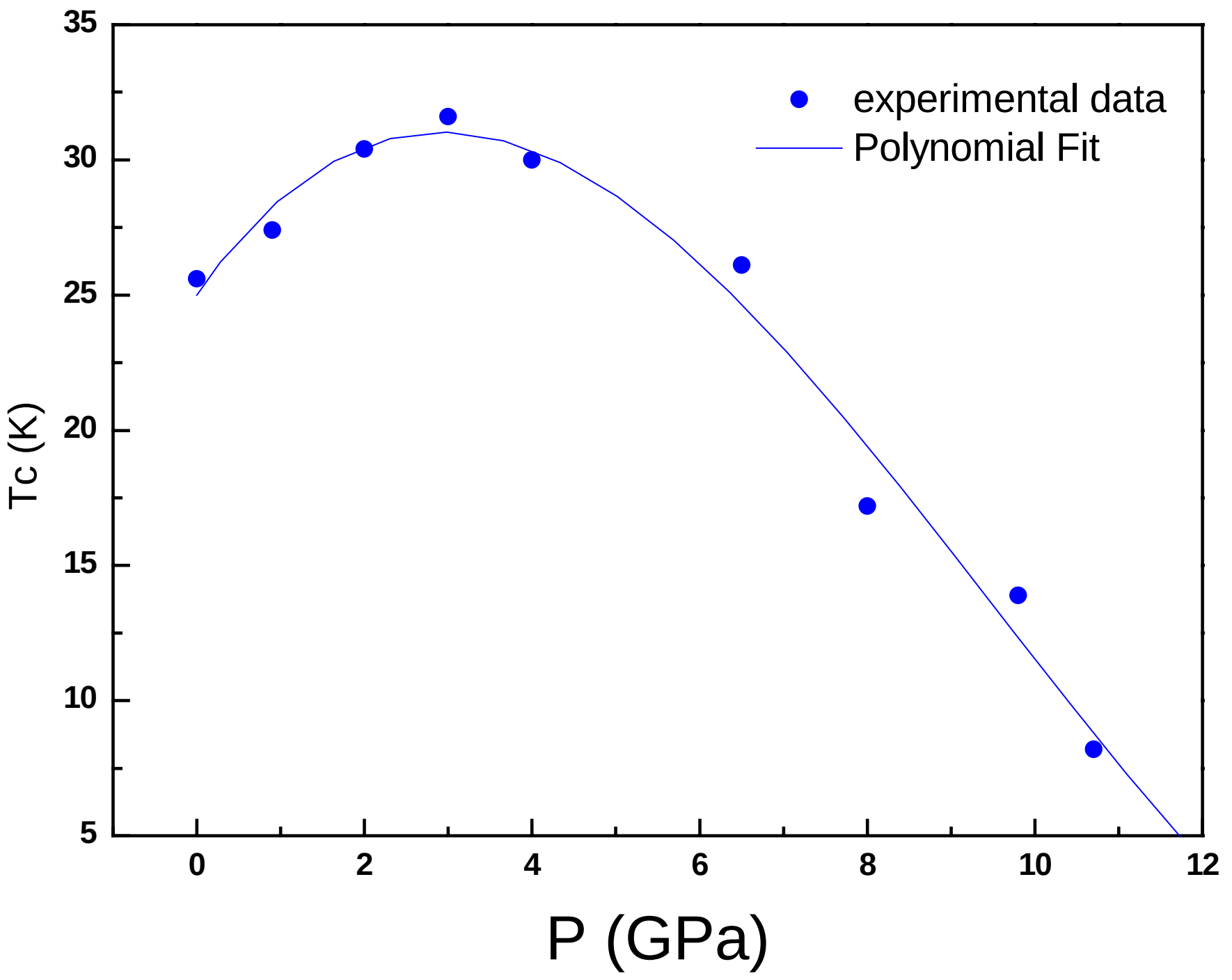

Fig.4

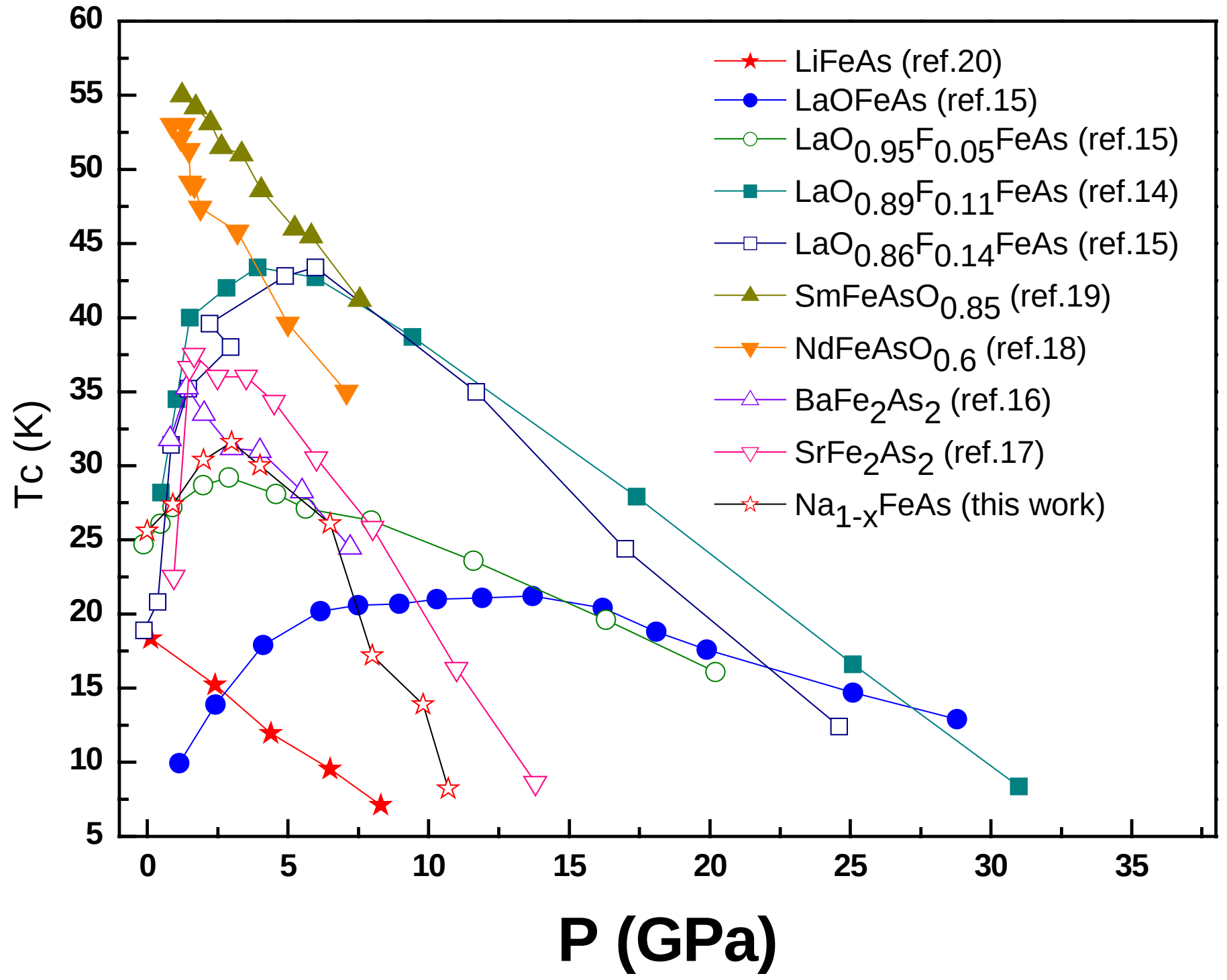

Fig.5